\documentclass[a4paper,11pt]{iopart}
\usepackage{graphicx}
\usepackage{cite}
\usepackage{color}
\usepackage{amsfonts}
\begin{document}
\title{Entropy production of a bound nonequilibrium interface}
\author{A. C. Barato$^1$ and H. Hinrichsen$^2$}
\address{$^1$The Abdus Salam International Centre for Theoretical Physics\\
		 34014 Trieste, Italy\\
         $^2$Universit\"at W\"urzburg,
	 Fakult\"at f\"ur Physik und Astronomie\\
         D-97074 W\"urzburg, Germany}

\ead{acardoso@ictp.it}

\begin{abstract}
We study the entropy production of a microscopic model for nonequilibrium wetting. We show that, in contrast to the equilibrium case, a bound interface in a nonequilibrium steady state produces entropy. Interestingly, in some regions of the phase diagram a bound interface produces more entropy than a free interface. Moreover, by solving exactly a four-site system, we find that the first derivative of the entropy production with respect to the control parameter displays a discontinuity at the critical point of the wetting transition. 
\end{abstract}

%==========================================================================
\section{Introduction}
%==========================================================================

For systems in thermal equilibrium the probability distribution of states is given by the Boltzmann-Gibbs measure, allowing one to calculate various macroscopic observables. Although such a theoretical concept is not available for systems out of equilibrium, it is nevertheless possible to make certain general statements in the nonequilibrium case. The most important statements of this kind are fluctuation relations~\cite{evans93,evans94,gallavotti95,Jarzynski,Crooks,kurchan98,lebowitz99,maes99,Jia1,seifert05,andrieux07,harris07,kurchan07,seifert08}, which constrain the probability distribution of fluctuating entropy along a given stochastic trajectory of microstates. A less general result, analogous to the second law of thermodynamics, is that the average entropy production is non-negative~\cite{schnakenberg76}. 

The average entropy production in the stationary state is a signature of nonequilibrium because it is zero if and only if detailed balance holds. Recent studies on the average entropy production explore its relation with the nonequilibrium steady state measure \cite{zia07,chetrite10,dorosz09,platini11,dorosz11} and its behavior at the critical point in systems with nonequilibrium phase transitions\cite{gaspard04,tome,andrae10,mario1,mario2}. Interestingly, in rather different models the entropy production was found to peak near the critical point. Moreover, the first derivative of the entropy production with respect to the control parameter was found to diverge at the critical point for the majority-vote model \cite{tome} and in a nonequilibrium Ising model in contact with two reservoirs \cite{mario1,mario2}. However, it remains unclear whether these are general properties of nonequilibrium models with continuous phase transitions. 

In this paper we address these questions by studying nonequilibrium wetting transitions~\cite{munoz04,santos04,barato10} which are known to exhibit a rich critical behavior out of equilibrium. Such transitions occur in models of growing interfaces belonging to the Kardar-Parisi-Zhang \cite{KPZ} universality class in the presence of a hard wall. The wetting transition is controlled by a parameter such as the growth rate, depending on which the interface either detaches or stays bounded to the wall.

As a representative of this class of models we investigate the so-called restricted solid-on-solid (RSOS) model with a hard wall at zero height~\cite{hinrichsen97}, extending a previous work where where we studied the entropy production of the free interface without a wall~\cite{chetrite10}. The presence of a hard wall changes the scenario completely: a bound phase arises in the region of the phase diagram where the free interface velocity would be negative. We show that a bound interface can produce entropy if detailed balance is not fulfilled and, moreover, in some regions of the phase diagram a stationary bound interface has a higher entropy production than a moving free interface. 

We also calculate the entropy production exactly for a four-site system. Interestingly, concerning entropy production, this small system displays the same qualitative behavior as an infinite one. As our main result, we show that in the non-equilibrium case, where detailed balance is violated, the first derivative of the entropy production  displays a discontinuity with respect to the control parameter at the critical point.     

The paper is organized as follows. In the next section we first recall the definition of the model and briefly explain the wetting transition. In Sec. 3 we define entropy production, discuss its interpretation in the RSOS model and show numerical simulation results. Our main findings come in Sec. 4, where we present the exact calculations for a four-site system. Finally, after concluding in Sec. 5, some intuitive ideas about entropy production are discussed in the appendix. 

%==========================================================================
\section{The model and its phase diagram}
%==========================================================================

%=================================================
\begin{figure}
\centering\includegraphics[width=100mm]{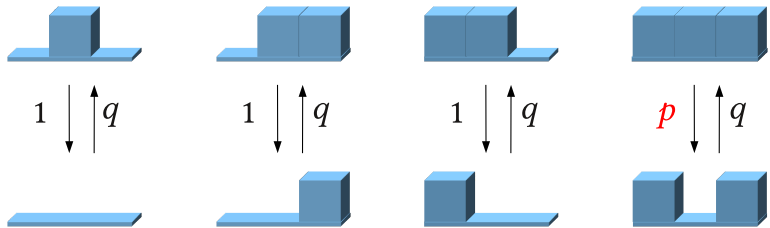}
\caption{Transition rates for the RSOS model.}
\label{fig1}
\end{figure}
%=================================================

The RSOS model studied in this paper is defined on a one-dimensional lattice with $L$ sites and periodic boundary conditions. The interface configuration is characterized by a height profile $\{h\}=(h_1,h_2,\ldots,h_L)$, where $h_i\in\mathbb{Z}$ is the height at site $i$. The interface evolves random-sequentially by spontaneous deposition and evaporation of particles with certain rates which are shown in Fig.~\ref{fig1}. Moreover, the dynamics is constrained by the restriction $|h_{i+1}-h_i|=0,1$, which introduces an effective surface tension. The model is controlled by two parameters, namely, the growth rate $q$ and the rate for evaporation from the middle of plateaus $p$, while the evaporation rate from edges is set to 1. As usual, the probability $P_{h_1,\ldots,h_L}(t)$ to find the system at time $t$ in the configuration $\{h\}$ evolves according to the master equation. 

In this paper we are primarily interested in the interface velocity and the entropy production. These quantities can be expressed in terms of the three-site probability distribution
\begin{equation}
P_{h_{i-1},h_{i},h_{i+1}}(t)\;=\;\sum_{h_1,\ldots,h_{i-2},h_{i+2},\ldots,h_L=0}^\infty P_{h_1,\ldots,h_L}(t)\,,
\label{3site}
\end{equation}
which is obtained by integrating out all other height variables. For periodic boundary conditions this distribution is translationally invariant and thus independent of $i$. From the dynamical rules in Fig \ref{fig1}, one can easily see that the interface velocity can be written in the form   
\begin{eqnarray}
\quad v &=&  q\sum_h\bigl(P_{h,h,h}+P_{h+1,h,h}+P_{h,h,h+1}+P_{h+1,h,h+1}\bigr)\nonumber\\
\qquad &&   -p\sum_hP_{h,h,h}-\sum_h\bigl(P_{h-1,h,h}+P_{h,h,h-1}+P_{h-1,h,h-1}\big)\,,
\label{defvelo}
\end{eqnarray}
where we suppressed the argument $t$ for the sake of readability. 

%=================================================
\begin{figure}
\centering\includegraphics[width=100mm]{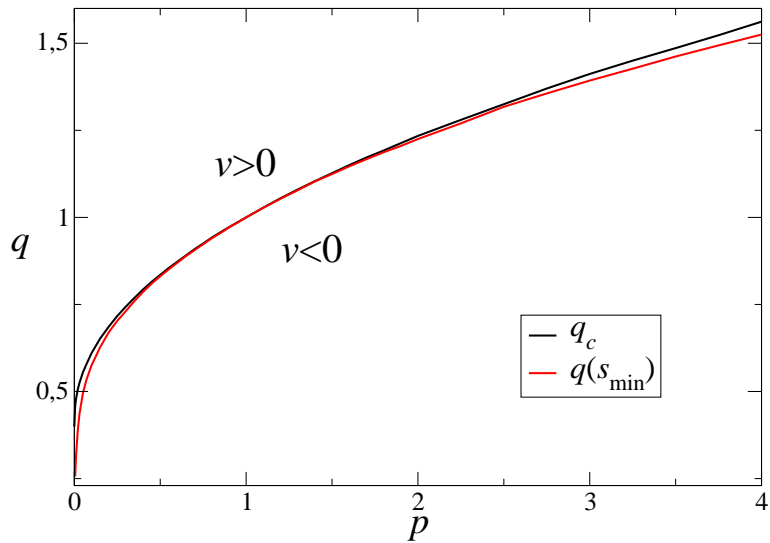}
\caption{Phase diagram of the RSOS model obtained from numerical simulations. Along the black line the asymptotic interface velocity is zero. The red line indicates the minimum of the free interface entropy production as a function of $q$ for a given $p$.}
\label{fig2}
\end{figure}
%=================================================

Depending on the control parameters $q$ and $p$ the interface of an infinite system will propagate (after an initial transient) with a constant positive or negative velocity~$v$. This defines two regions in the phase diagram (see  Fig.~\ref{fig2}), which are separated by the black line where $v=0$. For given $p$ we denote corresponding value of $q$ as $q_c=q_c(p)$. 

Let us now assume that the growth process takes place on top of an inert substrate by adding a hard wall at zero height. Such a wall can be introduced by forbidding evaporation at height $h=0$, meaning that negative heights become inaccessible~\cite{hinrichsen97,hlmp03}. Obviously, this modification does not change the behavior above the black line $q>q_c$, where the interface detaches from the wall and grows linearly in time with positive velocity. However, for $q\le q_c$ the situation changes completely: Instead of a negative velocity, the interface is now pinned to the wall and thus its velocity is zero. Therefore, the model with a wall exhibits a wetting transition at $q_c$ from a bound phase for $q\le q_c$ to a moving phase for $q>q_c$.   

A natural order parameter of the wetting transition is the density of contact points, where the interface touches the wall. In terms of this order parameter, the wetting transition displays a very rich critical behavior, which has been the main focus of past research in the context of nonequilibrium wetting \cite{munoz04,santos04,barato10}. In the present study, however, we are not interested in the contact point density, instead we will focus on the behavior of the entropy production. 

%==========================================================================
\section{Entropy production}
%==========================================================================

\subsection{Definition of Entropy production}
%--------------------------------------------
A continuous-time Markov process, as taking place in the RSOS the model, is defined by a space $\Omega$ of configurations $c\in\Omega$ and rates $w_{c\to c'}$ for spontaneous transitions from configuration $c$ to configuration $c'$. Denoting by $P(c,t)$ the probability of being in configuration $c$ at time $t$, the average entropy production in the environment (medium) is defined by~\cite{seifert05,schnakenberg76} (see Appendix)
\begin{equation}
S(t)= \sum_{c,c'}P(c,t)w_{c\to c'}\ln\frac{w_{c\to c'}}{w_{c'\to c}}\,.
\label{defentropy}
\end{equation}
Obviously, this expression requires all transitions to be reversible, i.e.
\begin{equation}
w_{c\to c'}\neq 0 \quad \Leftrightarrow \quad w_{c'\to c}\neq 0\,,
\end{equation}
meaning that the parameter $p$ in the RSOS model has to be positive. Of particular interest in this paper will be the average entropy production in the stationary state denoted by $S$, which is given by the above formula with the stationary measure $P(c)=\lim_{t\to\infty}P(c,t)$. Clearly, if detailed balance holds, i.e. $P(c)w_{c\to c'}=P(c')w_{c'\to c}$, the entropy production is zero. Because of the second law, $S$ is expected to be positive in a nonequilibrium stationary state.

\subsection{Entropy production in the RSOS model with a wall}
%--------------------------------------------

Using the three-site probability distribution (\ref{3site}) and  (\ref{defentropy}), the mean entropy production for the RSOS model  per site $s=S/L$  can be expressed as      
\begin{eqnarray}
s &=&\,\,\,\, \ln q \sum_h\bigl(P_{h,h,h}+P_{h+1,h,h}+P_{h,h,h+1}\bigr) \nonumber \\
  &&  -\ln q\sum_h\bigl(P_{h-1,h,h}+P_{h,h,h-1}+P_{h-1,h,h-1}\bigr) \nonumber \\
 &&   + \ln \frac{q}{p} \sum_hP_{h,h,h} -\ln \frac{q}{p} \sum_h P_{h,h,h}.
\label{entropyRSOS}
\end{eqnarray}
This expression is valid for the free interface, where the sums run over all integer heights, as well as for the bound case, where the summation is restricted to $h\geq 0$. As we divided by the lattice size, $s$ is an intensive quantity in $L$ so that we can conveniently carry out the limit $L \to \infty$.

For $p=1$ and $q<1$ the stationary bound state of the RSOS model with a wall is given by a Boltzmann-Gibbs-like exponential measure~\cite{hinrichsen97,hlmp03}
\begin{equation}
P(h_1,h_2,\ldots,h_L)\;\propto\; \chi{\{h\}} \prod_{i=1}^{L}q^{h_i},
\label{transfer}
\end{equation} 
where $\chi{\{h\}}$ is $1$ if the heights are non-negative and the configuration obeys the RSOS constraint and zero otherwise. In this situation the dynamics is known to obey detailed balance and hence the entropy production is zero. For $p=1$ and $q>1$, however, the model is in the moving phase so that detailed balance is not fulfilled. In this case one can show that the entropy production is positive and linearly related to the velocity by $s= v\ln q $.

%=================================================
\begin{figure}
\centering\includegraphics[width=160mm]{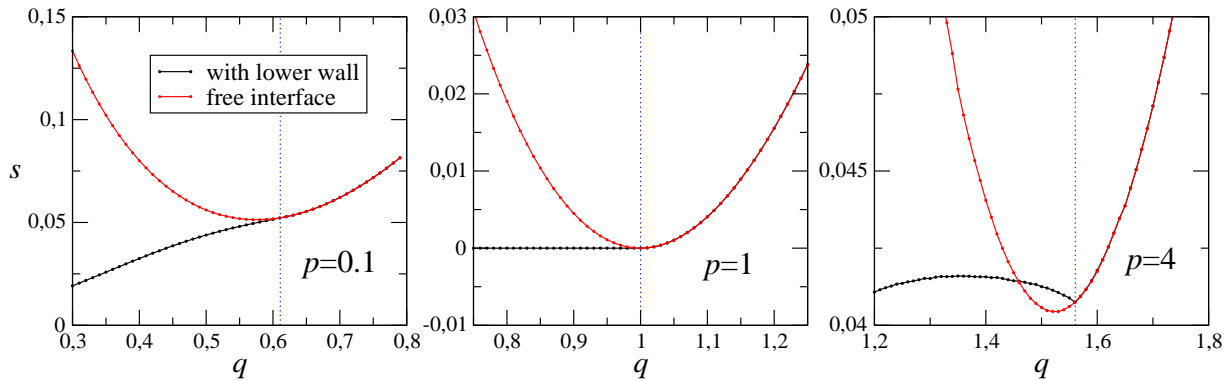}
\caption{Entropy production $s$ of the RSOS model for $p=0.1$ (left), $p=1$ (middle), and $p=4$ (right) with and without wall. The vertical dotted line indicates the transition point $q_c$. The simulations were done with system size $L=1024$.}
\label{fig3}	
\end{figure}
%=================================================

As shown in the Appendix, for $p\neq 1$, even in the bound phase, the system is out of equilibrium because different microscopic configurations are associated with different chemical potentials. Fig.~\ref{fig3} shows the entropy production of the RSOS model with and without the wall as a function of $q$ for various values of $p$. As expected, both entropies become identical at the critical point and coincide in the moving phase. Keeping $p$ fixed and varying $q$ the entropy production assumes its minimum at a certain value $q(s_{min})$. As can be seen in Fig.~\ref{fig3} and Fig.~\ref{fig2}, for $p\neq 1$ the location of this minimum is smaller than the critical point $q_c$. Interestingly, for $p=4$ the entropy production of the bound interface reaches a local maximum below the critical point. Moreover, there is an extended region where it is larger than the entropy production of a free interface. This means that the presence of the wall, which suppresses the flux of particles between the system and the reservoir, \textit{increases} the entropy production in the environment.

%==========================================================================
\section{Exact results for small system size}
%==========================================================================

The left and the central panels of Fig.~\ref{fig3} suggest that the entropy production of the interface with the wall (black curves) varies
smoothly at the critical point for $p\leq 1$ while the right panel indicates a discontinuous derivative for $p>1$. However, the numerical simulations are not accurate enough to confirm or disprove this conjecture. 

As a way out, let us consider \textit{finite} systems. In contrast to usual phase transitions, which exist only in infinite systems, the RSOS model with a wall exhibits a wetting transition even if the system size is finite. Although the critical behavior of such a finite RSOS model is completely different from the usual critical behavior in the limit $L\to \infty$~\cite{barato11,munoz}, we observe that entropy production is remarkably robust and that even very small systems display essentially the same qualitative behavior of this quantity as infinite ones.

\begin{figure}
\centering\includegraphics[width=100mm]{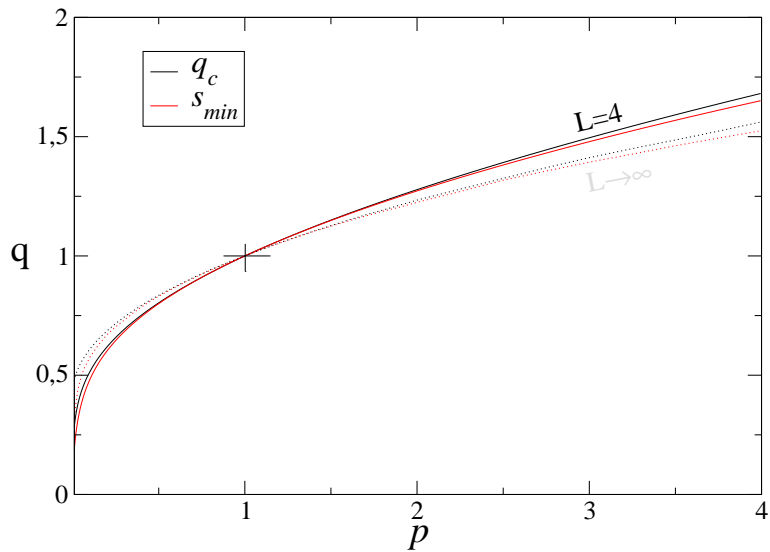}
\caption{Exact phase diagram for $L=4$. The dashed lines show the corresponding numerical data for $L\to\infty$ taken from Fig.~\ref{fig2}.}
\label{fig4}
\end{figure}

The smallest non-trivial system size $L=3$ is not appropriate for the present study because in this case detailed balance always hold in the bound phase. Therefore, we will focus on the case $L=4$, which was solved exactly in Ref.~\cite{barato11}. In this case we have (apart from vertical shifts) $19$ possible interface configurations. Because of translational invariance, only six of the 19 associated probabilities are independent, denoted as
\begin{eqnarray}
x_h &=& P_{h,h,h,h}\nonumber\\
y_h &=& P_{h,h,h,h+1}=P_{h,h,h+1,h}=P_{h,h+1,h,h}=P_{h+1,h,h,h}\nonumber\\
z_{h-1}&=& P_{h-1,h,h,h}=P_{h,h-1,h,h}=P_{h,h,h-1,h}=P_{h,h,h,h-1}\nonumber\\
u_h &=& P_{h,h+1,h+1,h}=P_{h,h,h+1,h+1}=P_{h+1,h,h,h+1}=P_{h+1,h+1,h,h}\nonumber\\
v_h &=& P_{h,h+1,h,h+1}=P_{h+1,h,h+1,h}\nonumber\\
w_h &=& P_{h,h+1,h,h-1}=P_{h-1,h,h+1,h}=P_{h,h-1,h,h+1}=P_{h+1,h,h-1,h}\,.
\label{def4}
\end{eqnarray}
For a free interface the RSOS model can be mapped onto charges jumping on a ring by introducing charge variables $\sigma_i=h_{i+1}-h_i=0,\pm 1$ \cite{neergaard97}. Therefore, the problem of calculating the stationary measure of the RSOS model with $L=4$ sites reduces to the problem of finding the eigenvector associated with the zero eigenvalue of a $6\times 6$ matrix \cite{chetrite10}. Following this procedure we can compute the entropy production (\ref{entropyRSOS}) and the velocity  (\ref{defvelo}) for the free interface,
\begin{eqnarray}
\fl s^{\rm (free)}(q,p)=\frac{16q^5+8q^4(p+1)+2q^2(q+p)(p-1)-2p(3+p)(2q+1+p)}{N(q,p)}\ln q
\label{entropyfree} \\
\fl\qquad\qquad\quad+\frac{8q^5+4q^4(p+1)-2q^2(q+p)(p-1)-2p(1+p)(2q+1+p)}{N(q,p)}\ln \frac{q}{p} \nonumber\\
\fl v^{\rm (free)}(q,p)=\frac{4 (1 + p + 2 q) (3 q^4-p (2 + p) )}{N(q,p)}\,,
\end{eqnarray}
where the denominator is given by
\begin{eqnarray}
N(q,p)&=&6 p^3+10 p^2 q^2+26 p^2 q+22 p^2+48 p q^3+59 p q^2\\
\nonumber &&+58 p q+17 p+47 q^4+47 q^3+27 q^2+11 q+2\,.
\end{eqnarray}
Setting $ v^{\rm (free)}(q,p)=0$ gives the critical line \cite{barato11}
\begin{equation}
p= -1+\sqrt{1+3q_c^4}.
\label{criticalline}
\end{equation}
In Fig. \ref{fig4} we plot the critical line together with the line where the entropy production (\ref{entropyfree}) is minimal. As in the case $L\to \infty$, the line of minimal entropy production lies always slightly below the critical line and touches it at the equilibrium point $p=1$. 

%=================================================
%=================================================

\begin{figure}
\centering\includegraphics[width=100mm]{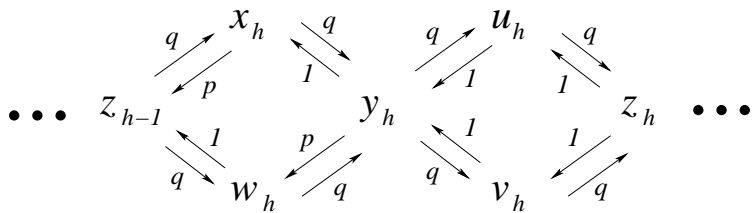}
\caption{Transition rates for $L=4$ in terms of the six variables defined in Eq.~(\ref{def4}).}
\label{fig5}
\end{figure}

Let us now turn to the entropy production of a \textit{bound} interface. As the wall breaks translational invariance in height direction, it is no longer possible to map the problem onto charges jumping on a ring. Therefore, we use a different method which was introduced in Ref.~\cite{barato11} to compute the density of contact points. Applied to the entropy production this method works as follows: In terms of the six variables (\ref{def4}) the dynamic rules in a system with $L=4$ sites form a network of transitions which is shown in Fig.~\ref{fig5}. With the help of this figure it is easy to see that the corresponding master equation is given by
\begin{eqnarray}
\dot x_h &=&  4y_h+4q z_{h-1}-4[q+p(1-\delta_{h,0})]x_h\nonumber\\
\dot y_h &=&  qx_h+2u_h+v_h+qw_h-[1+p(1-\delta_{h,0})+3q]y_h\nonumber\\
\dot z_h &=&  px_{h+1}+qv_h+2qu_h+w_{h+1}-(p+2+2q)z_h\nonumber\\
\dot u_h &=&  2qy_h+2z_h-2(1+q)u_h\nonumber\\
\dot v_h &=&  2qy_h+2pz_h-2(p+q)v_h\nonumber\\
\dot w_h &=&  qz_{h-1}+py_h-(q+1)w_h,
\label{masterL4}
\end{eqnarray}
where $w_0=0$ and all variables for negative heights are zero. Note that the six variables represent 19 different configurations, which is taken into account by certain multiplicities in the equations. For example, the factor 4 in term $4y_h$ in the first equation comes from the fact that there are four different height profiles represented by $y_h$ and they all go to $x_h$ by means of an evaporation event at rate $1$. Moreover, note that translational invariance in height direction is broken by the height-dependent terms on the right-hand side. For example, the terms $(1-\delta_{h,0})$ come from the fact that in the presence of a wall evaporation is forbidden at zero height.

%=================================================
%=================================================

The stationary solution of the master equation (~\ref{masterL4}) reads
\begin{eqnarray}
\fl x_h= Z^{-1}\bigg(\frac{3q^4}{p(2+p)}\bigg)^h,\qquad y_h= qx_h,\qquad z_h= \frac{3q^3}{2+p}x_h,\qquad w_h=px_h, \nonumber\\
\fl u_h= \bigg(\frac{q^2}{q+1}+\frac{3q^3}{(q+1)(2+p)}\bigg)x_h,\qquad v_h= \bigg(\frac{q^2}{q+1}+\frac{3pq^3}{(q+1)(2+p)}\bigg)x_h, 
\end{eqnarray}
where $Z$ is a normalization factor. This factor can be determined by computing the normalization $\sum_{h=0}^{\infty}P_h=1$ of the single-site probability distribution $P_h$, which in terms of the six variables is given by
\begin{eqnarray}
P_h &=& x_h+3y_h+y_{h-1}+z_h+3z_{h-1}+2u_h\nonumber\\
&&+2u_{h-1}+v_h+v_{h-1}+2w_h+w_{h-1}+w_{h+1}
\end{eqnarray}
so that
\begin{equation}
Z^{-1}= \frac{2 p + p^2 - 3 q^4}{p (2 + p + 8 q + 4 p q + 12 q^2 + 6 p q^2 + 12 q^3 + 12 q^4)}\,.
\end{equation}
Finally, expressing the three-site probability distribution (\ref{3site}) in terms of the four-site probability distribution (\ref{def4}) and using formula (\ref{entropyRSOS}), we obtain 
\begin{equation}
\fl s^{\rm (bound)}(q,p)= \frac{2 (p-1) q^3}{(1 + q) (2 + p + 8 q + 4 p q + 12 q^2 + 6 p q^2 + 12 q^3 + 12 q^4)}\,\ln p\,,
\label{sbound}
\end{equation}
which is valid in the stationary bound state below the critical line (\ref{criticalline}). Therefore, the entropy production of the model with a wall at zero height can be expressed as
\begin{equation}
s(q,p)= s^{\rm (bound)}(q,p)\theta(q-q_c)+s^{\rm (free)}(q,p)\theta(q_c-q)\,,
\label{sL4}
\end{equation}
where $\theta(x)$ is the Heaviside step function, $q_c$ is given by (\ref{criticalline}), $s^{\rm (free)}(q,p)$ by ({\ref{entropyfree}}) and $s^{\rm (bound)}(q,p)$ by (\ref{sbound}).

%=================================================
\begin{figure}
\centering
\includegraphics[width=150mm]{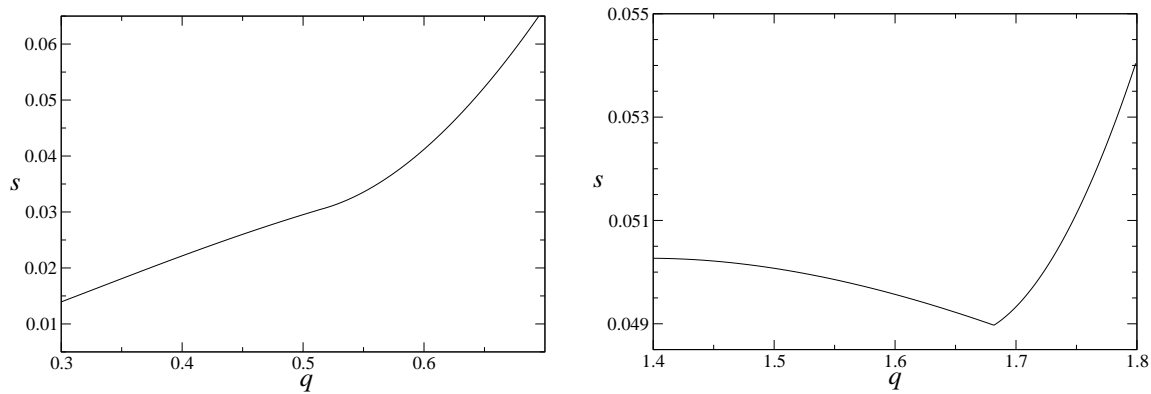}
\caption{The entropy production (\ref{sL4}) $s(q)$ for $p=0.1$ (left) and $p=4$ (right).}
\label{fig6}	
\end{figure}
%=================================================

\begin{figure}
\centering
\includegraphics[width=150mm]{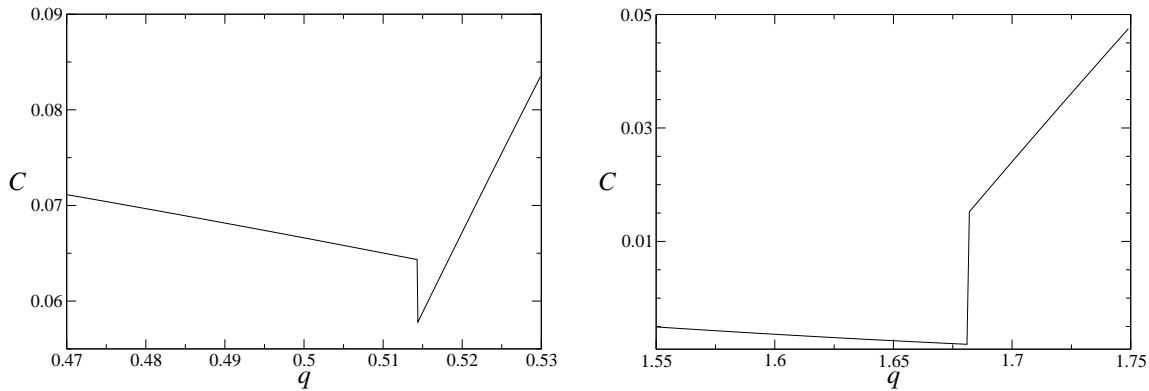}
\caption{The first derivative of the entropy production $C(q,p)$ for $p=0.1$ (left) and $p=4$ (right).}
\label{fig7}	
\end{figure}

In Fig. \ref{fig6} we plot the entropy production (\ref{sL4}) as a function of $q$ for two different values of $p$. It is clear that it is continuous at the critical point, which can be verified analytically by checking that (\ref{entropyfree}) and (\ref{sbound}) become identical along the critical line. Contrarily, the first derivative with respect to $q$, denoted as 
\begin{equation}
C(q,p)= \frac{\partial}{\partial q}s(q,p).
\end{equation} 
is found to be discontinuous at the critical point for $p \neq 1$. The explicit expression for this derivative is too cumbersome to be written down. However, plotting the derivative in Fig. \ref{fig7} for $p=0.1$ and $p=4$ one can clearly see a jump at the critical point. This is different from the results obtained in other nonequilibrium models \cite{tome,mario1,mario2}, where the first derivative was found to diverge at criticality. 

The only exception is the case of detailed balance $p=1$. Here the entropy production of the free interface varies as $\ln q (q-1)$ and vanishes at the critical point $q_c=1$ while the entropy production of a bound interface vanishes for any $q<q_c$. Therefore, the first derivative is continuous for $p=1$, independent of the system size. We therefore conclude that the observed discontinuity is a characteristic property of nonequilibrium in the present model. Morevoer, we conjecture that the size of the jump is a measure of how far the system is driven away from equilibrium. 

%=================================================
%=================================================

%==========================================================================
\section{Conclusion}
%==========================================================================

In this paper we have studied entropy production in the RSOS model with a wall which displays a nonequilibrium wetting transition. We have shown that in nonequilibrium a bound interface produces entropy due to an imbalance in the chemical potential associated with different local microscopic configurations. Surprisingly, we find that in some regions of the phase diagram a nonequilibrium bound interface produces more entropy than a moving one.

We have solved the model exactly for the case $L=4$, allowing for a precise analysis of the entropy production at the critical point. The main result of the paper is that the first derivative of the entropy production with respect to the control parameter $q$ displays a discontinuity at the critical point. Furthermore, such a discontinuity occurs only under nonequilibrium conditions, whereas for $p=1$, where detailed balance is fulfilled in the bound phase, the first derivative is found to be continuous. 

The entropy production was found to peak near the critical point in other nonequilibrium phase transitions~\cite{gaspard04,tome,andrae10,mario1,mario2} and its first derivative was found to diverge at the critical point \cite{tome,mario1,mario2}. In this paper we showed that in nonequilibrium wetting transitions its first derivative displays a discontinuity at criticality. It would be interesting to further investigate the critical behavior of the entropy production in other systems out of equilibrium, it might be that the entropy production plays an important role in classifying different nonequilibrium phase transitions.

%==========================================================================
\appendix
\section{Motivation of the formula for entropy production}
%==========================================================================

Formula (\ref{defentropy})  gives a mathematical quantity that, in principle, can be calculated for any Markov process. The name entropy production of the external medium can be justified as follows. Consider a system ($S$) in contact with a thermal reservoir ($R$) at temperature $T$ with the stationary measure $P(c)\propto \exp[- E(c)/T]$. For such a system the detailed balance condition reads
\begin{equation}
\ln\frac{ w_{c\to c'}}{w_{c'\to c}}= -\frac{1}{T}(E(c')-E(c)).
\label{DB}
\end{equation}
If the stationary system jumps from configuration $c\to c'$, its energy changes by $\Delta E_{S}(c\to c')= E(c')-E(c)$ while the energy of the external reservoir changes oppositely by $\Delta E_{R}(c\to c')=-\Delta E_{s}(c\to c')$. As the temperature of the reservoir is constant, the detailed balance condition (\ref{DB}) implies that
\begin{equation}
\Delta S = \frac{\Delta E_{R}(c\to c')}{T}=\ln\frac{ w_{c\to c'}}{w_{c'\to c}}.
\label{Ereservoir}
\end{equation}
In order to make the formula for entropy production plausible, let us now consider a fictitious system which is randomly switching between two thermal reservoirs at different temperatures $T_1$ and $T_2$. Such situation is described by a Markov process with transition rates $w_{c'\to c}=w^{(1)}_{c'\to c}+w^{(2)}_{c'\to c}$, where the rates $w^{(1)}_{c'\to c}$ and $w^{(2)}_{c'\to c}$ are the equilibrium rates for each of the two reservoirs.

In general, such a system is out of equilibrium and its stationary measure is not known. Let us assume that the system is in its nonequilibrium stationary state, undergoing a sequence of $N$ transitions $c_0\to c_1\to\ldots\to c_N$.  Using formula (\ref{Ereservoir}) the variation of the energy in the reservoirs $\nu=1,2$ can be written as
\begin{equation}
\label{DER}
\Delta E_R^{(\nu)}= T_\nu\sum_{i=1}^{N}\delta_{\nu,\nu_i}\ln\frac{ w^{(\nu_i)}_{c_{i-1}\to c_{i}}}{w^{(\nu_i)}_{c_{i}\to c_{i-1}}},
\end{equation}
where $\nu_i=1,2$ labels the reservoir that the system was connected to during the jump $c_{i-1}\to c_i$. The total change of the entropy in the external medium (consisting of the two reservoirs) is then given by the sum of the energy differences in each reservoir divided by its respective temperature, i.e.
\begin{equation}
\Delta S\;=\;\frac{\Delta E_R^{(1)}}{T_1}+\frac{\Delta E_R^{(2)}}{T_2}= \sum_{i=1}^{N}\ln\frac{ w^{(\nu_i)}_{c_{i-1}\to c_{i}}}{w^{(\nu_i)}_{c_{i}\to c_{i-1}}} \,.
\end{equation}
Finally, fixing a time interval $\Delta t$ and averaging $\Delta S/\Delta t$ over of all possible trajectories we obtain the formula for entropy production (\ref{defentropy}) in the stationary state. Note that these arguments rely on the assumption that the two reservoirs are almost equilibrated, which is reflected by using (\ref{Ereservoir}) in (\ref{DER}). This assumption is discussed in more detail in Ref.~\cite{HGJ11}.

The above argument can be used to explain the positive entropy production of the RSOS model in the stationary state for $p\neq 1$. To this end let us consider the growth process as a system in contact with a reservoir of particles, where the chemical potential for the deposition and evaporation of particles depends on the local microscopic configuration. If the interface separates two different phases $A$ and $B$, deposition and evaporation events can be interpreted as reactions $B\to A$  and $A\to B$, respectively. Let us take the equilibrium case $p=1$ and denote by $\mu_A$ and $\mu_B$ the chemical potentials of  $A$ and $B$, respectively. Hence if $N$ particles are deposited, the energy in the external reservoir will change by $\Delta E_R= N\Delta\mu= N(\mu_B-\mu_A)$, where $q= \e^{\Delta \mu}$. 

Now let us assume that the system is coupled to two reservoirs, namely one reservoir with chemical potential difference $\Delta\mu_1$ for deposition and evaporation on flat parts and at the edges and a second reservoir for evaporation from plateaus with chemical potential difference $\Delta\mu_2$. Assuming the reservoirs to be almost equilibrated we may use the detailed balance condition (\ref{DB}) to identify $q=e^{\Delta\mu_1}$ and $q/p=e^{\Delta\mu_2}$. When the two chemical potentials are different ($p\neq 1$) the system is out of equilibrium and there is an average flux of energy between the system and the reservoir even in the bound phase. The flux components (growth velocities) related to the chemical potential differences $\Delta\mu_1$ and $\Delta\mu_2$ are 
\begin{eqnarray}
v_1&=& q\sum_h\bigl(P_{h,h,h}+P_{h+1,h,h}+P_{h,h,h+1}-P_{h-1,h,h}-P_{h,h,h-1}-P_{h-1,h,h-1}\bigl),\nonumber\\
v_2&=& q\sum_hP_{h+1,h,h+1}-p\sum_hP_{h,h,h}.
\end{eqnarray}
Therefore, $v= v_1+v_2$ and $s= \Delta\mu_1v_1+\Delta\mu_2v_2$. In the bound phase we have $v_1=-v_2$ and the entropy production is $s= v_1(\Delta\mu_1-\Delta\mu_2)$. Hence, out of equilibrium, the imbalance of the chemical potentials for different microscopic configurations causes the entropy production in the bound phase to be positive. 

%==========================================================================
\section*{References}
%==========================================================================

\providecommand{\newblock}{}

\end{document}